\documentclass[a4paper,twocolumn]{article}%[a4paper,twocolumn,10pt]

\usepackage{amssymb}
\usepackage{graphicx}
\usepackage{amsmath,amsfonts,amsthm,bm} 
\usepackage{mathtools}
\usepackage{times}
\usepackage{bm}
\usepackage{url}
\usepackage{multirow}
\usepackage{tabularx, ragged2e, booktabs}
\newcolumntype{Y}{>{\centering\arraybackslash}X}
\usepackage{adjustbox}
\usepackage{cprotect} % for verb in figure caption https://tex.stackexchange.com/questions/8810/how-to-include-verbatim-in-a-figure-caption

\DeclareMathOperator{\EX}{\textrm{E}}
\DeclareMathOperator{\Var}{\textrm{Var}}
\DeclareMathOperator{\diff}{\textrm{d}}

\usepackage[update,prepend]{epstopdf} % But it doesn't work if eps is in another folder. Then compile from command window: pdflatex -shell-escape presentation.tex

\begin{document}

\title{An efficient algorithm to estimate the potential barrier height from noise-induced escape time data}

\author{\large Tam\'as B\'odai$^{1,2}$\\
$^1$Centre for the Mathematics of Planet Earth, University of Reading, UK\\ 
$^2$Department of Mathematics and Statistics, University of Reading, UK
} 

\maketitle

%\begin{abstract}

%\textit{Abstract.} We consider the problem of measuring a potential barrier in systems... [CONTINUE] ***NOT NEEDED FOR https://www.journals.elsevier.com/journal-of-computational-physics ALTERNATIVE JOURNALS: https://scicomp.stackexchange.com/questions/2868/alternatives-to-journal-of-computational-physics 

%\end{abstract}

\bigskip

%\section{Introduction}

It is a common phenomenon in nature and technology that a system under perturbations exits a regime of its usual dynamics~\cite{Kuehn_etal:2014,Dakos17546,Tang2017,FARANDA201426,Takács2008,doi:10.1063/1.5027718}. Often it is possible to define a potential function whereby a potential well can be associated with a usual or persistent dynamics, and a saddle of the potential adjacent to the potential well is a feature through which the exit takes place~\cite{Risken}. The potential difference between the bottom of the potential well and the saddle is often termed a potential barrier. The expected exit time then depends on the height of this potential barrier and the (small) noise strength. Therefore, knowing the potential barrier height is often of strong interest, because then one can predict -- for a given or applied noise strength -- the expected escape time. 

We develop an algorithm to determine the potential barrier height experimentally, provided that we have control over the noise strength. We are concerned with the situation when the experiment requires large resources of time or computational power, and wish to find a protocol that provides the best estimate in a given amount of time. We encountered such a situation when wanted to determine expected transition times to a cold climate for a noisy version of the climate model presented in~\cite{0951-7715-30-7-R32}.

We consider the rather generic situation when the dynamics is governed by the following Langevin stochastic differential equation (SDE):

\begin{equation}\label{eq:langevin}
 %dx = -\frac{dV}{dx}dt + \sigma dW
 \dot{\mathbf{x}} = \mathbf{F}(\mathbf{x}) + \sigma\mathbf{D}\bm{\xi}(t),
\end{equation}
$\mathbf{x},\mathbf{F},\bm{\xi}\in\mathbb{R}^n$, and the diffusion matrix $\mathbf{D}\in\mathbb{R}^{n\times n}$ is independent of $\mathbf{x}$, i.e., the white noise $\bm{\xi}$ is additive. The vector field $\mathbf{F}(\mathbf{x})$ is such that it realises the coexistence of multiple attractors (including the possibility of an attractor at infinity) and at least one nonattracting invariant set, often called a saddle set. The saddle set is embedded in the boundary of some basins of attraction. Based on a well-established theory due to Freidlin and Wentzell~\cite{FW:1984} the steady state probability distribution in the weak-noise limit, $\sigma\ll1$, can be written as

\begin{equation}\label{eq:stationary_distr}
 %W(\mathbf{x}) \sim Z(\mathbf{x})\exp(- \Phi(\mathbf{x})/\sigma^2),
  W(\mathbf{x}) \sim Z(\mathbf{x})\exp(-2\Phi(\mathbf{x})/\sigma^2),
\end{equation}
in which $\Phi(\mathbf{x})$ is called the nonequilibrium- or quasi-potential. In gradient systems where $\mathbf{F}(\mathbf{x})=-\nabla V(\mathbf{x})$ we have that  %$\Phi(\mathbf{x})=2V(\mathbf{x})$, 
$\Phi(\mathbf{x})=V(\mathbf{x})$,
provided that $\mathbf{D}=\mathbf{I}$. If $\mathbf{D}$ does depend on $\mathbf{x}$, then %(\ref{eq:stationary_distr}) does not apply; [12.07.18. WAS IT MY OWN CONVICTION OR I HAVE REFERENCE FOR THIS? WHY CAN THE PREFACTOR NOT BE A POWER LAW? I MEAN TO SAY THAT THE PROCESS WITH MULTIPLICATIVE NOISE (NO NEED FOR CAM -- RECALL JEROEN WEISS' PAPER) HAS A STATIONARY DISTR IN THE LARGE DEVIATION FORM. OR IS IT NOT? IS A POWER LAW TAIL NOT ALLOWED INDEED? CHECK FREDY B'S SLIDES, AND RECALL WHAT THE FUNNY CURLY EQUALITY SIGN MEANS!! ALSO, YING DOES CALCULATE THE QUASI-POTENTIAL FOR MULTIPLICATIVE NOISE TOO. IS THERE A CONDITION WHAT PROCESSES ARE ELIGIBLE FOR THE LARGE DEVIATION LAW?]
% 18.07.18. 
$W(\mathbf{x})$ might not satisfy a large deviation law $\lim_{\sigma\to0}\sigma^2\ln W(\mathbf{x})=-2\Phi(\mathbf{x})$. See e.g.~\cite{PhysRevE.96.032120} for an example of multiplicative noise where $\lim_{\sigma\to0}\sigma^2\ln W(\mathbf{x}=x)$ does not exists for some parameter setting and $W(x)$ has a fat tail.

The probability that a perturbed trajectory does not escape the basin of attraction over a time span of $t_t$ decays exponentially:

\begin{equation}\label{eq:tt_distr}
 P(t_t) \sim \frac{1}{\tau}\exp(-t_t/\tau).
\end{equation}
The approximation is in fact quite good already for times $t_t\approx \EX[t_t]=\tau$ or even smaller. %[***SHOW PICTURE?] 
The reciprocal of the expectation value $\tau$ can be written as an integral of the probability current through the basin boundary, whose leading component as $\sigma\rightarrow0$ comes from a point $\mathbf{x}_e$ where  $\Phi(\mathbf{x})$ is minimal on the boundary. The proportionality of the probability current to $W(\mathbf{x})$ leads~\cite{LT:2011,doi:10.1117/12.546995} to:

\begin{equation}\label{eq:tau}
 %\tau \propto \exp( \Delta\Phi/\sigma^2),
  \tau \propto \exp(2\Delta\Phi/\sigma^2),
\end{equation}
where 

\begin{equation}\label{eq:DPhi}
 \Delta\Phi=\Phi(\mathbf{x}_e)-\Phi(A)
\end{equation}
%$\Delta\Phi=\Phi(\mathbf{x}_e)-\Phi(A)$ 
is what we call the potential barrier height. Both the saddle and the attractor can be chaotic, in which cases $\Phi(\mathbf{x}_e)$ and $\Phi(A)$ have been shown~\cite{HAMM1994313,PhysRevLett.66.3089} to be constant over the saddle~\cite{HAMM1994313} % 02.08.18. Why did TT write in email that "a pot. konstans a repellorokon, ill a nyergek instabil sokasagan (nem az egesz nyergen)". First, the saddle is part of its unstable manifold, so if the pot is constant on the u manifold then it's constant over the saddle. However, surely it's not constant over the WHOLE unstab man, because that extends all the way to the attractor, where the pot is clearly a lot smaller.
and attractor~\cite{PhysRevLett.66.3089}, respectively.

Considering (\ref{eq:tau}), the expected transition times increase ``explosively'' as the noise strength $\sigma$ decreases. From the point of view of estimating $\Delta\Phi$, there seems to be a trade-off between an increasing accuracy of the estimation and an increasing demand of resources as $\sigma$ decreases. However, if we fix the amount of resources that we are willing to commit, then an increasing of accuracy is not guaranteed any more, because we can register fewer transitions as $\sigma$ decreases. On the other hand, increasing $\sigma$ beyond a point might not improve accuracy either for the following reason. We assume that for some $\sigma_0$ we can estimate $\tau=\tau_0$ arbitrarily accurately because a large number of transitions can be achieved inexpensively. We also assume that in this ``anchor point'' (\ref{eq:tau}) applies accurately: 

\begin{equation}\label{eq:tau2}
 \tau \approx \tau_0\exp(2\Delta\Phi(\sigma^{-2} - \sigma_0^{-2})), \quad \sigma<\sigma_0.
\end{equation}
Then, we can identify the accuracy of estimation by 

\begin{equation}
 \delta\Delta\Phi = \frac{\sqrt{\Var[\ln \bar{t}_t]}}{y},
\end{equation}
where we introduced $y=\sigma^{-2} - \sigma_0^{-2}$, and $\bar{t}_t=\frac{1}{N}\sum_{i=1}^Nt_t$ is our finite-$N$ estimate of $\tau$ for a fixed $\sigma$. Clearly, as $\sigma\rightarrow\sigma_0$ the inaccuracy explodes. That is, in the described setting of estimation (which is not the most generic one) there should exist an optimal value of $\sigma$. 

The sum of the exponentially distributed random variables, $N\bar{t}_t$, does in fact follow an Erlang distribution~\cite{EHP:2000}, and so:

\begin{equation}\label{eq:Erlang}
 P(\bar{t}_t) \sim \frac{1}{\tau^N}\frac{(N\bar{t}_t)^{N-1}}{(N-1)!}\exp(-N\bar{t}_t/\tau)N.
\end{equation}
Note that since $\EX[\bar{t}_t]=\EX[t_t]=\tau$, our estimator $\bar{t}_t$ is unbiased. Furthermore, $\Var[\bar{t}_t]=\Var[t_t]/N=\tau^2/N$ in accordance with the Central Limit Theorem. From (\ref{eq:Erlang}) it follows %[***SHOULD I TRY TO REPRODUCE THE RESULT FROM MATHEMATICA BY HAND?] 
that

\begin{equation}
 \Var[\ln \bar{t}_t] = \Psi^{(1)}(N),
\end{equation}
where $\Psi^{(1)}(N)$ is the first derivative of the digamma function~\cite{AS:1972}. We can make the interesting observation that $\Var[\ln \bar{t}_t]$ does not depend on $\tau$, only on $N$. Next, we make use of the approximation~\cite{AS:1972}

\begin{equation}\label{eq:approx_Psi}
 \Psi^{(1)}(N) \sim 1/N
\end{equation}
writing

\begin{equation}
 \delta\Delta\Phi \sim \sqrt{\frac{\tau_0}{T}}\frac{\exp(\Delta\Phi y)}{y},
\end{equation}
where we, first, assumed a certain fixed commitment of resources, which can be expressed simply by $T=N\tau$, and, second, made use of (\ref{eq:tau2}). We look for a $\sigma=\sigma^*$ or $y=y^*$ that minimizes $\delta\Delta\Phi$, for which we need to solve $\diff\delta\Delta\Phi/\diff y=0$, yielding our main result:

\begin{equation}\label{eq:opt}
 y^*=\Delta\Phi^{-1}.
\end{equation}
We can make the interesting observation that it is independent of $\tau_0$ and $T$, which we comment on shortly. $y^*$ depends only on $\Delta\Phi$ (in a very simple way), the unknown that we wanted to determine in the first place, and so the result can seem irrelevant to practice for the first sight. However, one can simply start out with an initial guess value, $\hat{\Delta\Phi}_0$, and iteratively update the estimate as $\hat{\Delta\Phi}_{i}$ by performing a maximum likelihood estimation (MLE)~\cite{Coles:2001} each time a new value of $t_{t,i}$ is acquired. This way, for the acquisition of $t_{t,i+1}$, one continues the experiment with an updated noise strength $y_{i+1}^*=\hat{\Delta\Phi}^{-1}_i$, $i=1,\dots,N$, according to (\ref{eq:opt}). The MLE of $\Delta\Phi$ is based on the probability distribution (\ref{eq:tt_distr}) jointly with (\ref{eq:tau2}). This is an analogous procedure to nonstationary extreme value statistics when one or more parameters of the extreme value distribution (EVD) is a function of a covariate that could depend on time. In our case $\tau$ and $\sigma$ correspond to the EVD parameter and covariate, respectively. We note that as $\sigma^*$ does not depend on $T$, at any time into the experiment (for large enough $N$, though, such that (\ref{eq:approx_Psi}) is a good approximation) our estimate of $\Delta\Phi$ is done most efficiently, and so we can revise our commitment, either stopping the experiment early or extending it. Next we demonstrate the use of our algorithm on two examples; in a single- as well as a multi-dimensional system. %[***IS THERE ANYTHING TO SAY ABOUT THE INDEPENDENCE OF $\tau_0$? SUCH AS: TRY WITH SOME ANCHOR WITH VERY LARGE NOISE, THEN TRY WITH ANOTHER ANCHOR, STILL ACCEPTABLE, AND SEE IF THE ANCHOR IS GOOD. THIS MIGHT STILL BE MORE EFFICIENT THAN HAVING EQUIDISTANT SIGMA SAMPLE VALUES -- THE USUAL WAY. PERHAPS THE REFEREE WILL ASK.]

\begin{figure}  %[t!]
    \begin{center}
        \begin{tabular}{cc}
            \includegraphics[width=\linewidth]{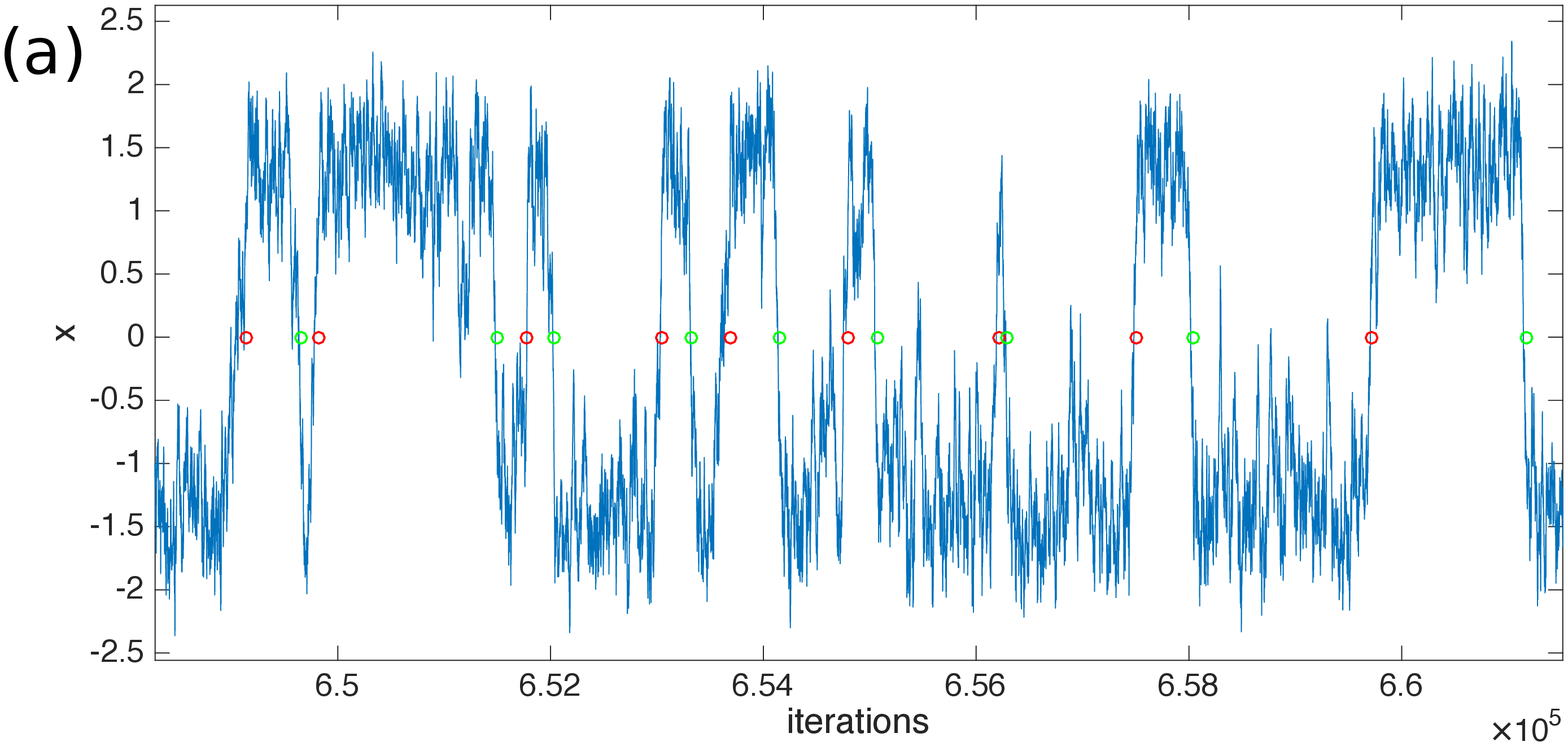} \\
            \includegraphics[width=\linewidth]{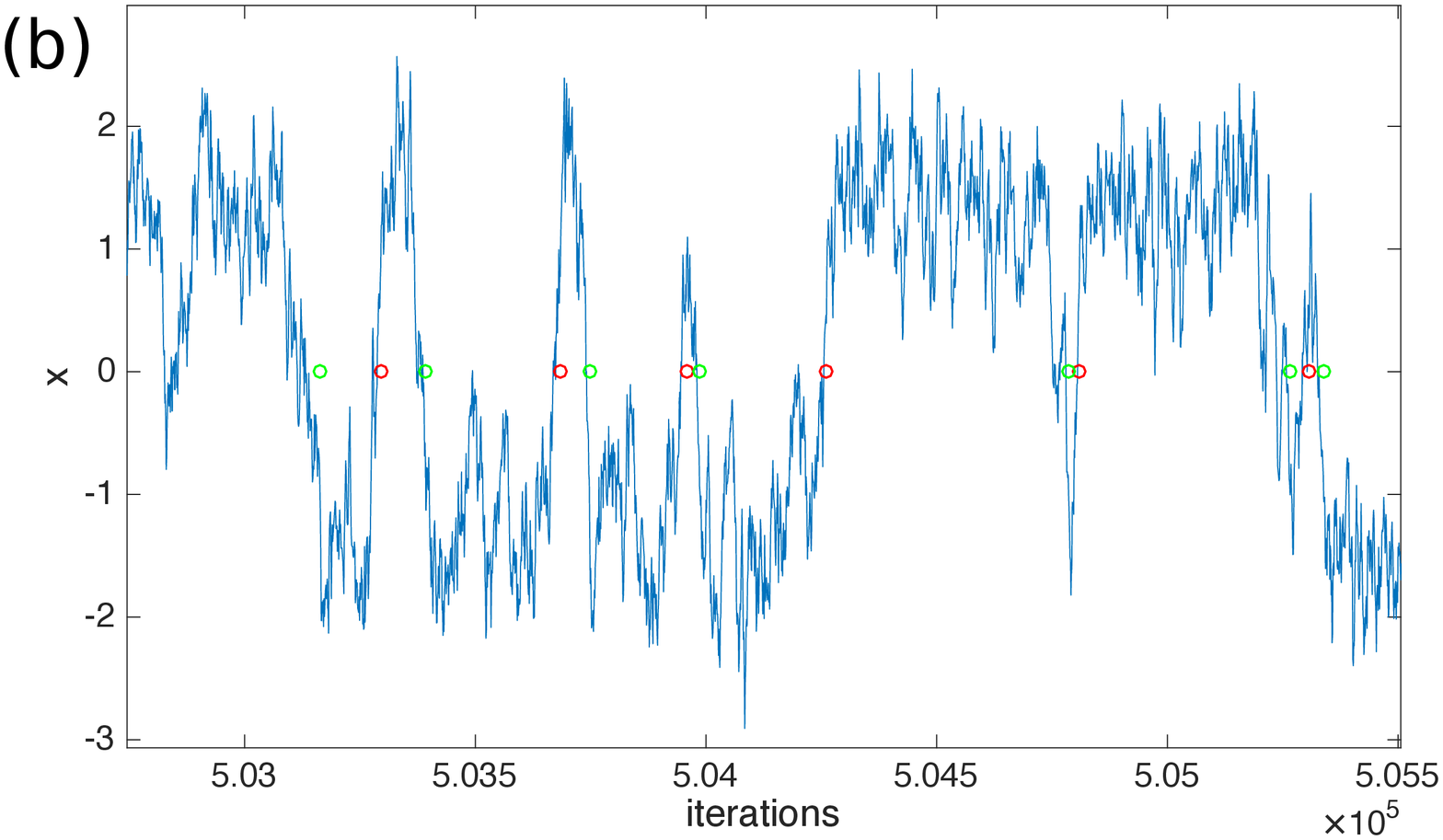} 
        \end{tabular}
        \caption{\label{fig:escape_time_direct_06} Numerical solution of (\ref{eq:sde_1d}). (a) $\sigma=1.0$, (b) $\sigma=1.55$. Red and green circle markers indicate transition times defined as a first crossing to the bottom of the upper (lower) potential well since a crossing to the lower (upper) well. 
        }
    \end{center}
\end{figure}

\begin{figure}  %[t!]
    \begin{center}
        \begin{tabular}{cc}
            \includegraphics[width=\linewidth]{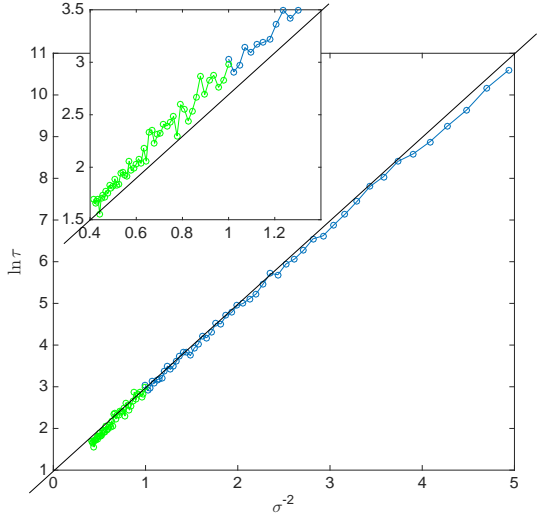} 
        \end{tabular}
        \caption{\label{fig:escape_time_direct_03} Demonstration of the validity of (\ref{eq:tau}) in (\ref{eq:sde_1d}). A straight line of slope $\Delta V=1$ is included in the diagram for reference. %$N=50???$  [***ALTHOUGH THE SCALING HOLDS HERE FOR RATHER LARGE $\sigma$, AND SO NO PROBLEM SEEMS TO BE WITH FINDING AN ANCHOR, THIS IS NOT THE CASE IN PUMA-GSEBM.]
        To estimate $\tau$ we averaged $N=200$ transition times each sample values of $\sigma$ corresponding to a circle marker. % I guess this is based on the escpate_time_direct.m script that i shared with VL in a dropbox. The escpate_time_direct.m i have now on my comp' is a much upgraded version and i played with the parameters.
        }
    \end{center}
\end{figure}

\begin{figure}  %[t!]
    \begin{center}
        \begin{tabular}{cc}
            \includegraphics[width=\linewidth]{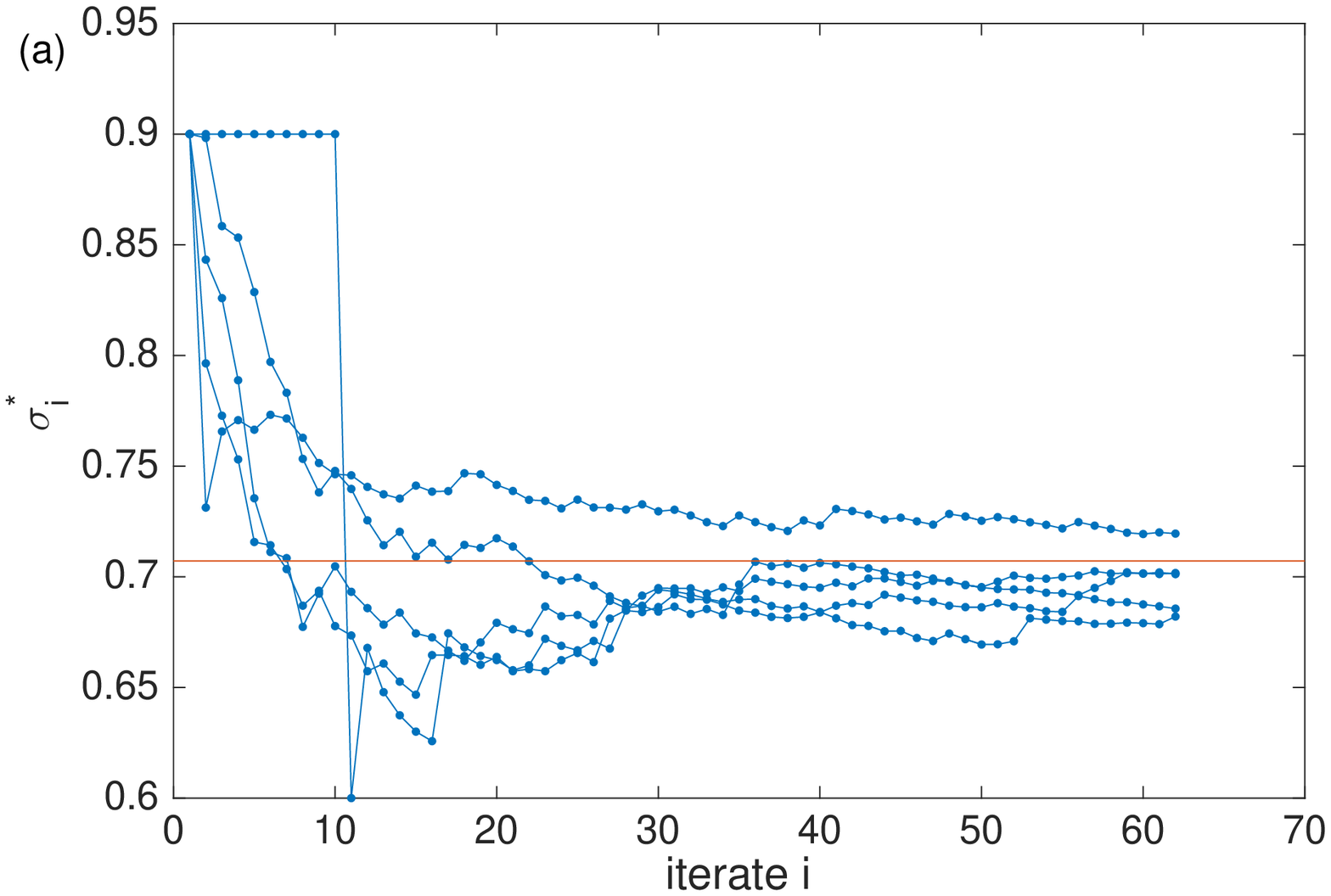} \\
            \includegraphics[width=\linewidth]{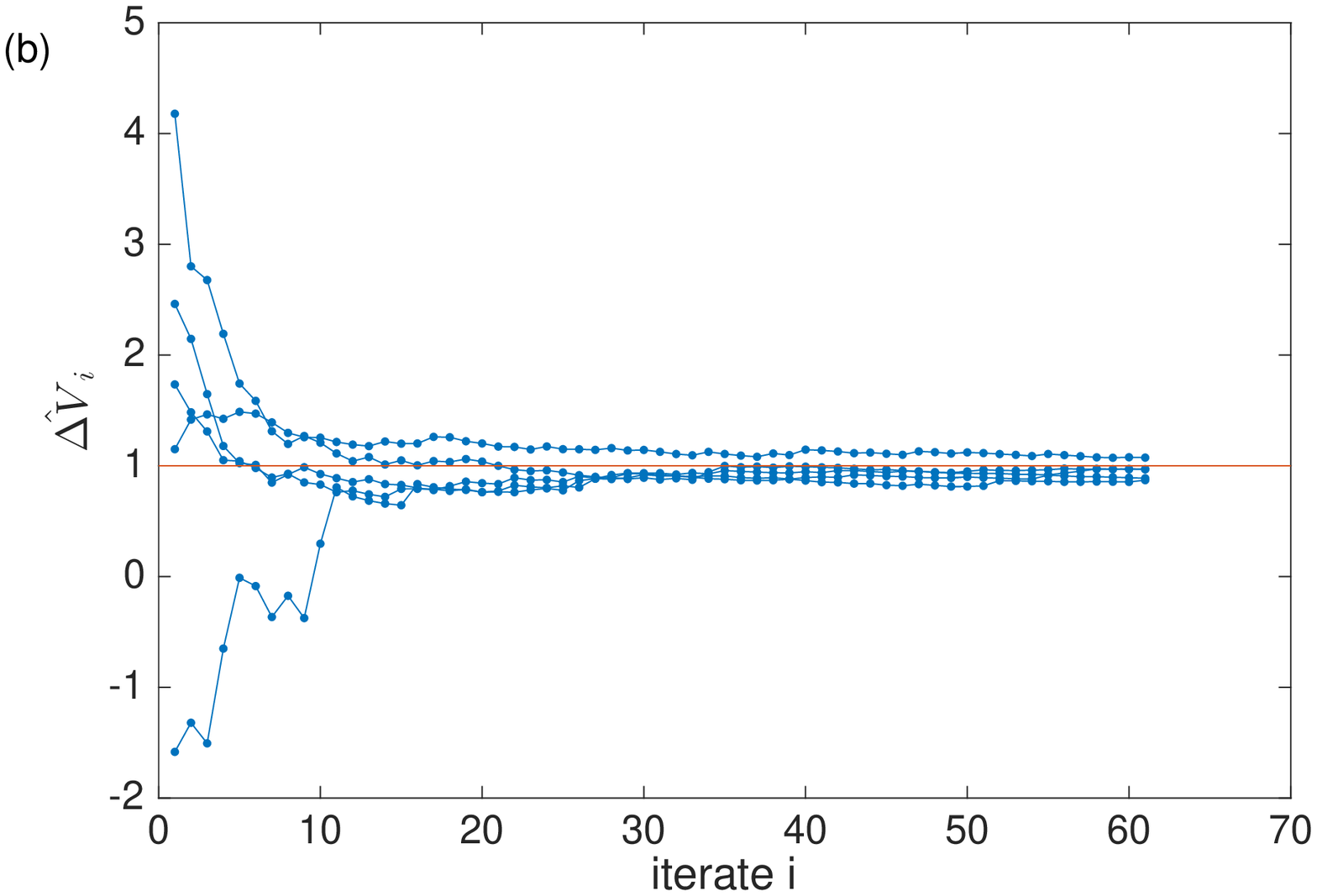} 
        \end{tabular}
        \caption{\label{fig:escape_time_direct_07} Proof of concept, I: convergence of estimates $\hat{\Delta\Phi}_i$. %Performance of the proposed algorithm. NOT REALLY THE PERFORMANCE, 
        The anchor $\tau_0$ was established with $\sigma_0=1$ using $N=N_0=400$. Five different realizations of the experiment are shown. The initial value for each was $\sigma_0^*=0.9<\sigma_0$. A ``safeguarding'' of the procedure is facilitated by overriding (\ref{eq:opt}) such that $\sigma_{i+1}^*=\sigma_0$ if $\hat{\Delta\Phi}_i<0$ and $\sigma_{i}^*=\sigma_{min}^*=0.6$ when (\ref{eq:opt}) dictates smaller.
        }
    \end{center}
\end{figure}

\begin{figure}  %[t!]
    \begin{center}
        \begin{tabular}{cc}
            \includegraphics[width=\linewidth]{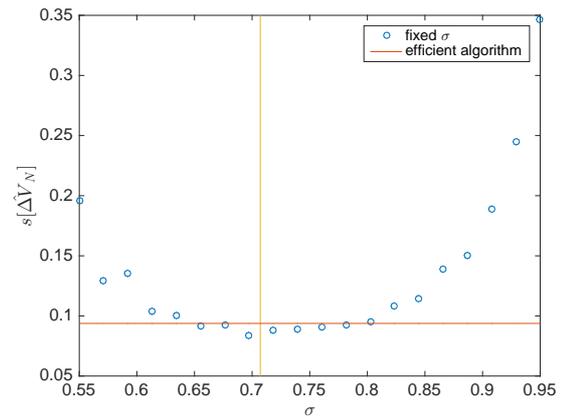} 
        \end{tabular}
        \caption{\label{fig:escape_time_direct_08_03} Proof of concept, II: sample standard deviation of 200 estimates $\hat{\Delta V}_N$. For the efficient algorithm we chose $N=30$, which implies (see the main text) the different $N$'s for the different fixed sample values of $\sigma$. The vertical line marks the prediction of (\ref{eq:opt}). 
        }
    \end{center}
\end{figure}

{\em Example 1: Overdamped particle in a symmetrical 1D double-well potential.} It is governed by the following SDE:

\begin{equation}\label{eq:sde_1d}
 %dx = -\frac{dV}{dx}dt + \sigma dW
 dx = -V'dt + \sigma dW.
\end{equation}
We specify our example as: $V=x^4/4-x^2$. The two minima are at $x_{\pm}=\pm\sqrt{2}$, and the local maximum in between is at $x_0=0$. These are fixed points of the deterministic case ($\sigma=0$). A numerical solution of the SDE (\ref{eq:sde_1d}) is obtained by using an Euler-Maruyama integrator~\cite{Kloeden_Platen} with a time step size of $h=0.02$. Examples are shown in Fig. \ref{fig:escape_time_direct_06}, indicating the regime behaviour with irregular transitions between the two regimes. %[***BUT ALSO THAT $x_0$ CAN BE PASSED WITHOUT A FULLY-FLEDGED TRANSITION.] 
The time series clearly evidence bimodal marginal distributions -- corresponding to the two regimes -- whose maxima, and the local minimum in between (not shown), are exactly at $x_{\pm}=\pm\sqrt{2}$ and $x_0=0$, respectively.
%the minima of the potential wells, and its local maxima in between the wells, respectively, due to (\ref{eq:stationary_distr}). These are the fixed points of the deterministic system ($\sigma=0$): $x_{\pm}=\pm\sqrt{2}$, $x_0=0$. 
With substituting these in to (\ref{eq:DPhi}) we obtain that $\Delta V=\Delta\Phi=1$. This shows up as the slope of the curve in Fig. \ref{fig:escape_time_direct_03}. % 16.07.18.
The green coloring indicates that (\ref{eq:tau}) is satisfied well even with so strong noise that the time spent in a regime is not so clear cut any more, as seen in Fig. \ref{fig:escape_time_direct_06} (a). The result of applying our algorithm is shown in Fig. \ref{fig:escape_time_direct_07}, indicating that it serves its purpose, and that the convergence is rather fast. Finally, Fig. \ref{fig:escape_time_direct_08_03} verifies the corner stone of the algorithm (\ref{eq:opt}), showing the sample standard deviation of a number of estimates. Results with the algorithm and different fixed sample values of $\sigma$ are shown in one diagram, indicating that the accuracy of estimate by our algorithm is just about the best accuracy achievable by the same amount of computation using the optimal fixed $\sigma$. Note that we chose $N=30$ for our algorithm, resulting in some computational time $T$, and then we realised $N=\lceil T/\tau(\sigma)\rceil$ transitions using the different fixed $\sigma$'s.

% [*** INCLUDE THIS LATER IF STILL WANT TO]
% \begin{equation}\label{eq:tau3}
%  \tau \sim \frac{1}{2\pi}\sqrt{V''(x_+)|V''(x_0)|}\exp(2\Delta V/\sigma^2)
% \end{equation}

%[EXAMPLE 2: GSEBM, BUT I USE ADDITIVE NOISE, I.E., DON'T PERTURB MU. ***WOULD IN THIS CASE BE POSSIBLE TO CALCULATE $\Delta \Phi$ BY A RELAXING TRAJECTORY? I DEMONSTRATED THIS FOR 1D, BUT CAN I DO IT IN HIGHER-D? I LEFT A COMMENT ON THIS IN {\verb FW_exercise_04_Phi.m} IN THIS REGARD! I DON'T UNDERSTAND NOW WHY I WROTE IT WORKS ONLY IN 1D. THE ONLY CONDITION FOR DUALITY SEEMS TO BE THAT $D=I$.]

\begin{figure}  %[t!]
    \begin{center}
        \begin{tabular}{c}
            \includegraphics[width=\linewidth]{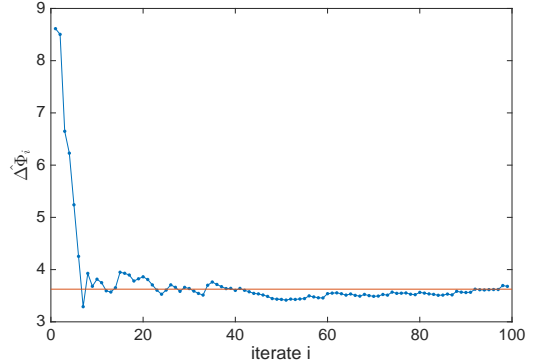} 
        \end{tabular}
        \cprotect\caption{\label{fig:simulate_noisy_GSEBM_04} Sames as Fig. \ref{fig:escape_time_direct_07} (b) but for the discretized GSEBM, $J=10$, and $\sigma_0=1.7$, $N_0=400$, $\sigma_0^*=1.5$, $\sigma_{min}^*=1.0$. We considered only warm-to-cold transitions given the present day solar strength $\mu=1$. The red line marks the potential difference $\Delta\Phi$ between the warm climate and the saddle. Note that here we measure time in $10^6$ seconds [Ms], as a change to~\cite{gsebm,BLLB:2014}. An expression for the potential functional $\Phi(T)$ of the PDE was given in~\cite{Ghil:1976}. However, we calculate $\Delta\Phi$ for the discretized system by an action-minimizing procedure~\cite{Tang2017}, using the computer code as a supplementary material to that paper. Time-discretization of the instanton was realised by 100 points over a span of 2000 [Ms]. Note that for the feasibility of the minimization it is crucial to provide symbolically the gradient of the action with respect to the displacement of the discrete sample points of the instanton. To be able to achieve this using the code of~\cite{Tang2017} the method of lines has to result in an explicit expression for $\dot{T_j}$. This could not be achieved by the sophisticated method implemented in Matlab's \verb|pdepe|, which is why we developed the discretization scheme (\ref{eq:discretization}).
        }
    \end{center}
\end{figure}

{\em Example 2: The Ghil-Sellers energy balance climate model (GSEBM).} One of the most striking facts about Earth's climate is its bistability: beside the relatively warm climate that we live in, under the present astronomical conditions a very cold climate featuring a fully glaciated so-called snowball Earth is also possible, and this state might have been experienced a number of times by Earth in the past few hundred million years~\cite{Hoffman_Schrag:2002}. Different hypotheses of transitioning from the warm to the cold climate and the other way round involve external forcings, but in principle it is possible that the climate system is transitive, at least in the warm-to-cold direction. This transitivity can be modeled by noise-induced transition, where the noise models some unresolved internal, say, atmospheric and/or oceanic dynamics. Without a requirement for physical realism, we consider additive noise perturbations of the Ghil-Sellers model~\cite{Ghil:1976} written for the long time average surface air temperature $T(\phi,t)$ as a function of latitude $\phi\in[-\pi/2,\pi/2]$ or $x=2\phi/\pi\in[-1,1]$. The deterministic GSEBM stands in the form of a diffusive heat equation:
% If i were pressed to put down the noise term in the eq., how would i do it in the case of a PDE??? I cannot even say properly that, beside the noise being additive, $D=I$, because that can be said only in terms of the ODE.
\begin{equation}\label{eq:gsebm}
  \begin{split}
    C(x)\partial_tT(t,x)= \mu Q(x)(1-\alpha(x,T))-O(T) + \\
      M(x)\partial_{x}[(D_1(x) + D_2(x)g(T))\partial_{x}T].
  \end{split}
\end{equation}
See~\cite{Ghil:1976,BLLB:2014} for the concrete form of the equation and the meaning of its terms, and~\cite{gsebm} for a numerical implementation. Unlike~\cite{gsebm} that uses Matlab's \verb|pdepe|, here we simulate the noise-perturbed GSEBM %(\ref{eq:gsebm}) commented out because it's not noise-perturbed but deterministic
by Matlab's \verb|simulate|. For this we discretize the eq. with respect to $T$ by the method of lines, converting the PDE in to an ODE, i.e., eq. (\ref{eq:langevin}). The particular difference schemes that we apply using a regular grid are: 
\begin{equation}\label{eq:discretization}
  \begin{split}
    \partial_x[D_1(x)\partial_xT]\approx [(T_{j+1}-T_{j})D_{1,j+1/2} - \\ (T_{j}-T_{j-1})D_{1,j-1/2}]/\Delta x^2, \\
    \partial_x[D_2(x)g(T)\partial_xT] = \partial_x[D_2(x)\partial_xG]\approx \\ [(G(T_{j+1})-G(T_{j}))D_{2,j+1/2} - \\ (G(T_{j})-G(T_{j-1}))D_{2,j-1/2}]/\Delta x^2,% This actually defines G
    %\nonumber
  \end{split}
\end{equation}
$j=1,\dots,J,$ where $T_j\approx T(x_j)$, $x_j=(j-1/2)\Delta x-1$, $\Delta x=2/J$, and $D_{1,j\pm1/2}=D_1(x_{j\pm1/2})$, $x_{j\pm1/2}=x_j\pm\Delta x/2$  %,$j=1,\dots,J$.
(see p. 1046 of~\cite{Press:2007} regarding the $x$-dependent diffusivity). The boundary conditions are eliminated by the method of reflection, setting $T_0=T_J$ and $T_{J+1}=T_1$. Such a grid deals effectively with the singularity of $M(x)$ at the poles, but the resulting ODE can be somewhat stiff. Fig. \ref{fig:simulate_noisy_GSEBM_04} shows that our algorithm works also in a multi-dimensional setting.

\section*{Acknowledgments}
I would like to thank Ying Tang for his extremely helpful support for using their code~\cite{Tang2017}, and Tam\'as T\'el for providing valuable feedback on a draft of the manuscript. I would like to acknowledge Valerio Lucarini for the many inspiring exchanges on the topic of critical transitions. This work received funding from the EU Blue-Action project (under grant No. 727852).

\bibliographystyle{unsrt} % 11.08.18. solution from: https://tex.stackexchange.com/questions/17354/sort-thebibliography-by-citation-order

%\bibliography{escape_time_algorithm}

\end{document}